\documentclass[aps,pre,twocolumn,showpacs,superscriptaddress,preprintnumbers]{revtex4}
\usepackage[utf8]{inputenc}
\usepackage{graphicx}
\usepackage{hyperref}
\usepackage{amsmath}
\usepackage{amssymb}
\usepackage{microtype}
\usepackage{natbib}
\usepackage{textcomp}
\usepackage[T1]{fontenc}
\usepackage{xspace}
\usepackage{subfigure}
\newcommand{\emg}{\textsc{emg}\oldstylenums{909}\xspace}
\makeatletter
\providecommand \@ifxundefined [1]{%
 \@ifx{#1\undefined}
}%
\providecommand \@ifnum [1]{%
 \ifnum #1\expandafter \@firstoftwo
 \else \expandafter \@secondoftwo
 \fi
}%
\providecommand \@ifx [1]{%
 \ifx #1\expandafter \@firstoftwo
 \else \expandafter \@secondoftwo
 \fi
}%
\providecommand \bibnamefont  [1]{#1}%
\providecommand \bibfnamefont [1]{#1}%
\providecommand \citenamefont [1]{#1}%
\providecommand \href@noop [0]{\@secondoftwo}%
\providecommand \href [0]{\begingroup \@sanitize@url \@href}%
\providecommand \@href[1]{\@@startlink{#1}\@@href}%
\providecommand \@@href[1]{\endgroup#1\@@endlink}%
\providecommand \@sanitize@url [0]{\catcode `\\12\catcode `\$12\catcode
  `\&12\catcode `\#12\catcode `\^12\catcode `\_12\catcode `\%12\relax}%
\providecommand \@@startlink[1]{}%
\providecommand \@@endlink[0]{}%
\providecommand \url  [0]{\begingroup\@sanitize@url \@url }%
\providecommand \@url [1]{\endgroup\@href {#1}{\urlprefix }}%
\providecommand \urlprefix  [0]{URL }%
\@ifxundefined {%
  \providecommand \doi  [0]{doi:\discretionary{}{}{}\begingroup
  \urlstyle{rm}\Url }%
}%
\providecommand \doibase [0]{http://dx.doi.org/}%
\providecommand \Doi [0]{\begingroup \@sanitize@url \@Doi }%
\providecommand \@Doi  [1]{\endgroup\@@startlink{\doibase#1}\@@Doi}%
\providecommand \@@Doi [1]{#1\@@endlink}%
\providecommand \selectlanguage [0]{\@gobble}%
\providecommand \bibinfo  [0]{\@secondoftwo}%
\providecommand \bibfield  [0]{\@secondoftwo}%
\providecommand \BibitemOpen [0]{}%
\providecommand \BibitemShut  [1]{\csname bibitem#1\endcsname}%

\begin{document}
\title{Magnetic traveling-stripe-forcing:\\ enhanced transport in the advent of the Rosensweig instability}
\author{Thomas Friedrich}
\affiliation{Experimentalphysik V, Universit\"at Bayreuth, D-95440 Bayreuth,Germany}
\author{Ingo Rehberg}
\affiliation{Experimentalphysik V, Universit\"at Bayreuth, D-95440 Bayreuth,Germany}
\author{Reinhard Richter}
\affiliation{Experimentalphysik V, Universit\"at Bayreuth, D-95440 Bayreuth,Germany}

\date{\today}

\begin{abstract}
A new kind of contactless pumping mechanism is realized in a layer of ferrofluid via a spatio-temporally modulated magnetic field. The resulting pressure gradient leads to a liquid ramp, which is measured by means of X-rays. The transport mechanism works best if a resonance of the surface waves with the driving is achieved. The behavior can be understood semi-quantitatively by considering the magnetically influenced dispersion relation of the fluid.
\end{abstract}

 \maketitle


\section{Introduction}
Ferrofluids are colloidal dispersions of magnetic nanoparticles, which show super-paramagnetic behavior \cite{rosensweig1985}. This offers the advantage to control and facilitate their flow by time-dependent magnetic fields. For example in alternating fields a viscosity reduction was predicted for Poiseuille flow \cite{shliomis1994} and thereafter experimentally confirmed \cite{bacri1995,zeuner1998}. More spectacularly, in rotating fields a spontaneous spin up of surface flow was observed in a beaker
\cite{moskowitz1967}, and quantitatively analyzed in a circular duct \cite{krauss2005}.

An intriguing alternative to rotating fields is to apply a spatio-temporally modulated field. Here one has to discern two transport mechanisms. One is ``ferrohydrodynamic pumping'' which was lately studied for a closed channel geometry \cite{mao2005,mao2006}. It relies on the phase shift between the external field and the magnetization and works best for frequencies in the range of the inverse Brownian relaxation time, i.e. in the kHz-regime. The second mechanism relies on traveling surface waves, driven by a traveling magnetic field. This was realized by an array of solenoids acting on a free surface of ferrofluid \cite{kikura1990}, and discussed in connection with the effect of peristaltic pumping \cite{zimmermann2004b}. As we will show in this article the driving caused by a free surface gains momentum if the traveling field is superimposed by a constant field oriented normally to the fluid layer.

In such a vertically oriented magnetic field, a plain layer of ferrofluid disintegrates into a hexagonal array of crests if a critical threshold $H_c$ is overcome -- the so called Rosensweig instability \cite{cowley1967}. Approaching $H_c$ from below, a vanishing of the phase velocity $v_p$ of surface waves was observed, and explained by the non-monotonic dispersion relation \cite{reimann2003}. In this regime, a traveling-stripe-forcing of the magnetic induction was predicted to excite a resonance of the wave amplitudes if the driving velocity coincides with $v_p$ \cite{bashtovoi1977ews}. This was recently verified by experiments \cite{beetz2008}. As we show in the following, this resonance is correlated with a maximal net force -- as indicated by a liquid ramp -- on the magnetic fluid.

\section{Experimental Setup and Procedure}
\label{sec:procedure}
\begin{figure}
\hspace{1.5cm}
\includegraphics[width=0.8\linewidth]{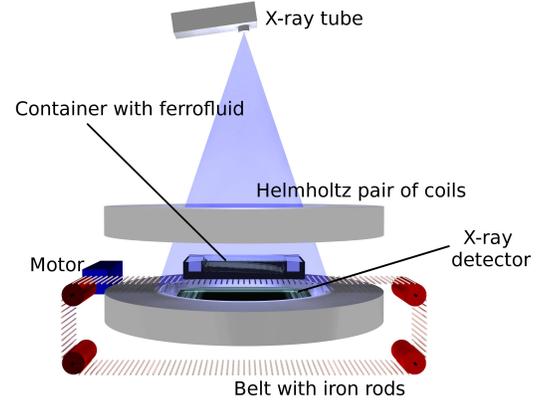}
\caption{Sketch of the experimental setup.}
\label{fig:setup}
\end{figure}

\begin{figure}
\includegraphics[scale=0.67]{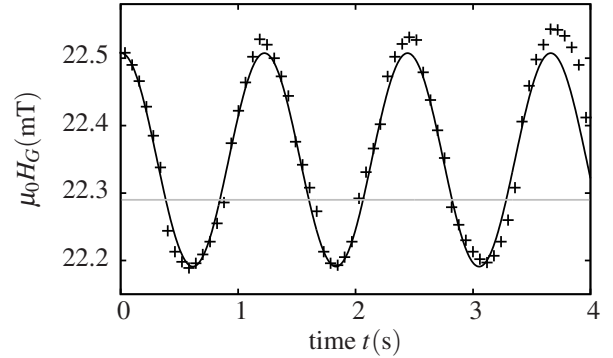}
\caption{The crosses indicate the magnetic field as a function of time. The horizontal gray line corresponds to the applied field of the Helmholtz coils $\mu_0 H_0$ as explained in the text. The black line represents a fitted harmonic function.}
\label{fig:magnetic.grid}
\end{figure}

Our setup is sketched in Fig.\,\ref{fig:setup}. A Helmholtz pair of coils provides an applied field $H_0$ along the vertical direction by means of a stabilized electrical current. In the center of the coils a container machined from Perspex\texttrademark\, is located. It contains a box shaped cavity with a length of $120\,\mathrm{mm}$, a width of $100\,\mathrm{mm}$ and a height of $25\,\mathrm{mm}$. This cavity is filled with $30\,\mathrm{ml}$ of ferrofluid \emg (Tab.\,\ref{tab:EMG909}). About $5\,\mathrm{mm}$ below the bottom of the cavity, a motor driven conveyor belt is located. It harbors a grid of rods made from welding wire ($\varnothing=2\,\mathrm{mm}$) with a spacing of $\lambda_G=(9.3 \pm 1)\,\mathrm{mm}$.

Radiation emitted by an X-ray tube mounted $1.5\,\mathrm{m}$ above the container is permeating the ferrofluidic layer and recorded by an X-ray detector (16 bit). The absorption pictures serve to reconstruct the surface topography of the liquid layer. The measurement of the height is calibrated by means of a ferrofluidic ramp \cite{richter2001,gollwitzer2007}. To eliminate the shadows cast by the rods we measure the time averaged absorption of X-rays with the moving belt included.

The belt leads to a spatial modulation of the magnetic field. The modulation of this field was measured within the empty cavity at one fixed position $2\,\mathrm{mm}$ above the ground of the container with the belt in motion. A result for a belt velocity $v_G$ of $0.75\,\frac{\mathrm{cm}}{\mathrm{s}}$ and an applied field of $\mu_0 H_0=22.29\,\mathrm{mT}$, which has been measured by the Hall probe in the absence of the grid, is depicted in Fig.\,\ref{fig:magnetic.grid}. The fit to a harmonic function demonstrates that the magnetic field can be modeled as a harmonic wave
\begin{equation}
H_G = H_{G,0} + \Delta H_G \sin{(\omega t - k_G x)}.
\label{eq:modulation}
\end{equation}
Here $H_{G,0}$ denotes the mean value and $\Delta H_{G,0}$ the amplitude of the modulated field. The frequency $\omega=2 \pi \frac{v_G}{\lambda_G}$\cite{beetz2008} is controlled by the motor (Mattke MDR-230). The wavenumber $k_G=\frac{2\pi}{\lambda_G}$ is set by the grid spacing, and has been chosen to be close to the critical wavenumber of the Rosensweig instability. The deviation between the value of the applied field $H_0$ as indicated by the gray line and the mean field $H_{G,0}$ is explained by a focusing of the magnetic field due to the grid.  
 
\begin{figure}
\includegraphics[scale=0.67]{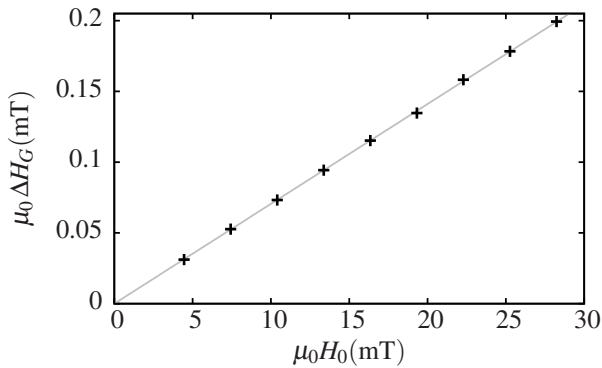}
\caption{Modulation amplitude vs.\ the induction impressed by the coils. The experimental data (black crosses) are recorded for a distance of $z=2\,\mathrm{mm}$ between the Hall probe and the ground of the container. Gray line: linear fit.}
\label{fig:deltaB_current}
\end{figure}
Fig.\,\ref{fig:deltaB_current} shows the modulation amplitude $\Delta H_G$ as a function of the applied magnetic field $H_0$. The fitted gray line indicates a linear dependence, thus demonstrating that the soft ferromagnetic material chosen for the rods is well suited for the experiments.

\begin{figure}
\includegraphics[scale=0.67]{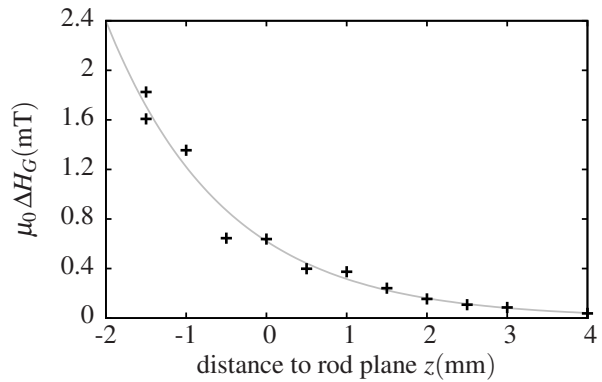}
\caption{$\Delta H_G$ measured for different distances $z$ at $\mu_0 H_0=22.29\,\mathrm{mT}$ (black crosses), gray line: exponential fit.}
\label{fig:deltaB_z}
\end{figure}

The modulation $\Delta H_G$ decreases with increasing distance from the rods. This is demonstrated experimentally in Fig.\,\ref{fig:deltaB_z}. The gray line is the fit to
\begin{equation}
 \Delta H_G(z)=\Delta H_G(0)\, e^{-z/\lambda_G},
\end{equation}
because we expect an exponential decay with the decay length given by the grid spacing $\lambda_G$ when assuming a two dimensional field. The fit yields $\mu_0\,\Delta H_G(0)=0.62\,\mathrm{mT}$ for the modulation at the bottom of the cavity.

\begin{figure}
\includegraphics[scale=0.67]{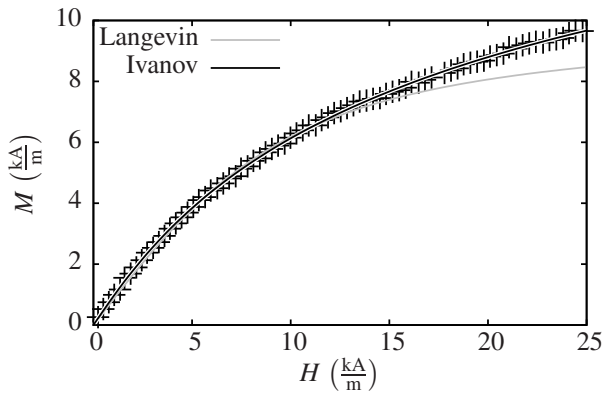}
\caption{Nonlinear magnetization curve $M(H)$. Black crosses: measured data, black line: fit using the Ivanov model \cite{ivanov2001}, gray line: fitted Langevin function.} \label{fig:magnetization.curve}
\end{figure}

The ferrofluid \emg is made of magnetite particles dispersed in kerosene. Its nonlinear magnetization curve is plotted in Fig.\,\ref{fig:magnetization.curve}. We measured $M(H)$ using a fluxmetric magnetometer consisting of a Helmholtz pair of sensing coils with $6800$ windings and a commercial integrator (Lakeshore Fluxmeter 480). The sample is held in a spherical cavity with a diameter of $12.4\,\mathrm{mm}$ in order to provide a homogeneous magnetic field inside the sample with a demagnetization factor of $\frac{1}{3}$. The gray line is a fit using the Langevin function \cite{rosensweig1985}
\begin{equation}
M = M_s \left( \coth{\left(\beta H \right)} - \frac{1}{\beta H}\right).
\label{eq:langevin}
\end{equation}
This function was derived for dilute monodisperse colloidal suspensions and shows visible deviations from the data. In contrast, the black line displays a fit of a model for dense polydisperse magnetic fluids, namely Eq.\,(32) of Ref.\,\cite{ivanov2001}. This model assumes a $\Gamma$- distribution for the particle diameter $d$
\begin{equation}
g(d) = \frac{1}{\Gamma(\alpha+1)d} \left(\frac{d}{d_0}\right)^\alpha \exp\left(-\frac{d}{d_0}\right) ,
\label{eq:gr_verteilung}
\end{equation}
where $d_0=1.4\,\mathrm{nm}$ and $\alpha=3.8$ are the obtained fit parameters, corresponding to a mean particle diameter $\bar{d}=6.6\,\mathrm{nm}$. The fit also provides the volume fraction of magnetite to be $4.0\%$.

Furthermore we characterize our ferrofluid by subjecting it to a homogeneous magnetic field, i.e.  we omit for this  measurement the driving belt. The plain layer becomes unstable at a critical field of $\mu_0 H_\mathrm{c,meas} = 21\,\mathrm{mT}$. Like in Refs.\,\cite{richter2005,gollwitzer2007} this value was obtained by fitting the measured spike amplitudes with a scaling law provided by theory \cite{friedrichs2001}.

\begin{figure}
\begin{center}
\begin{center}
(a)\hspace{0.14\columnwidth}
\includegraphics[width=0.74\linewidth]{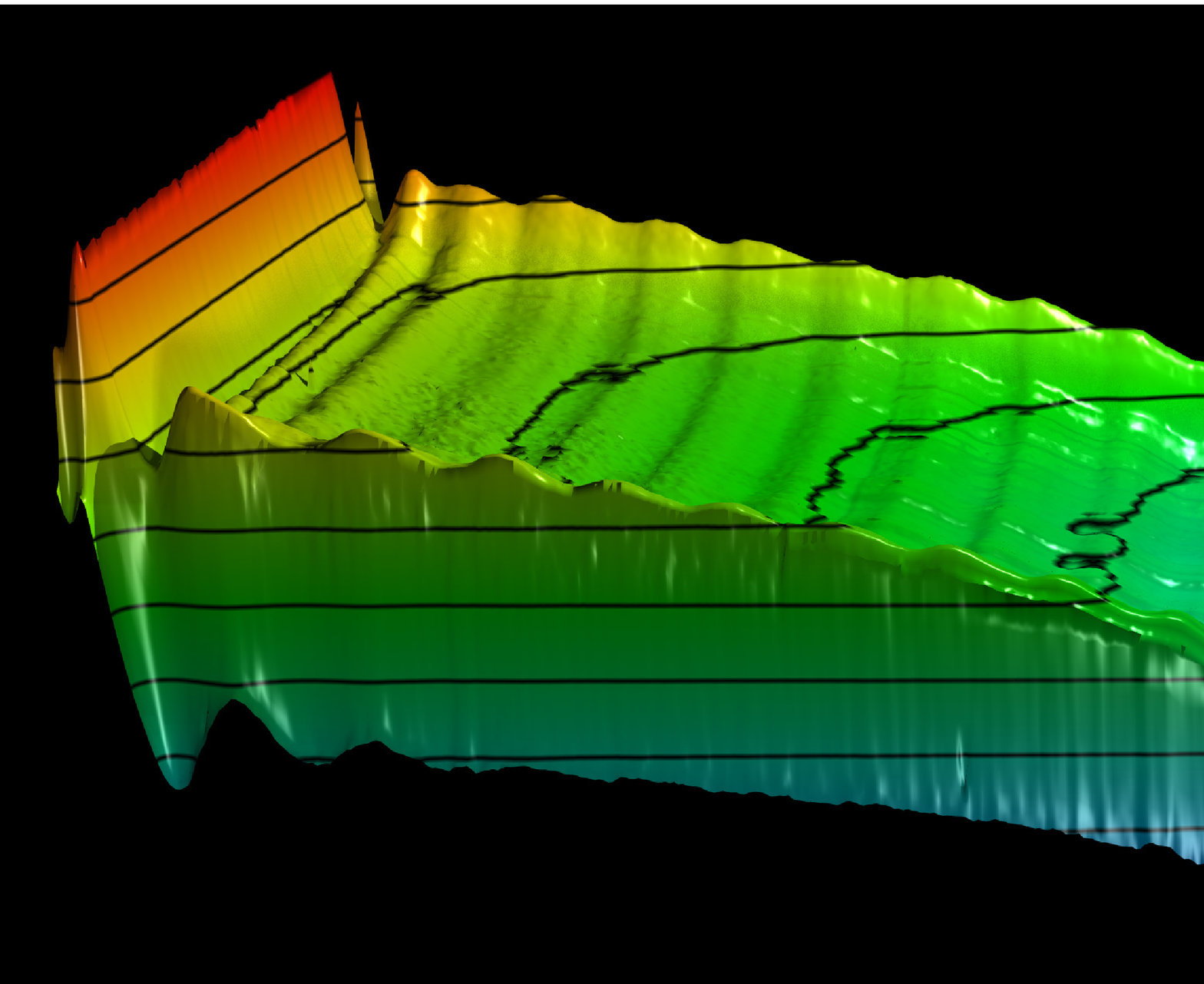}\\
(b)
\includegraphics[scale=0.67]{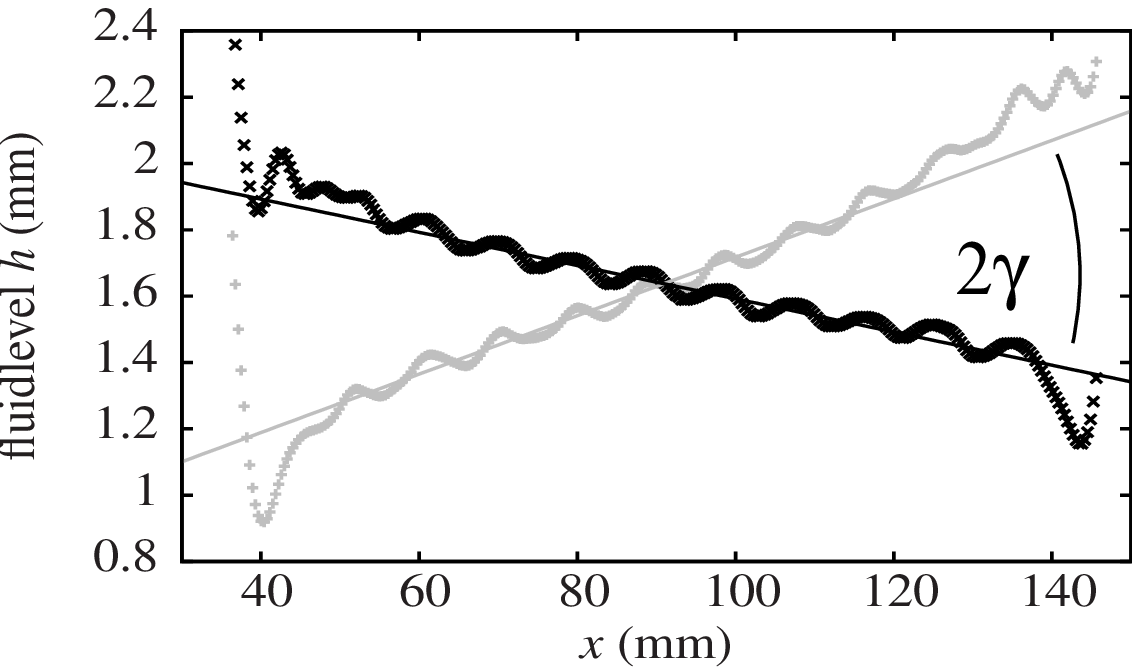}\\
\end{center}
\caption{A liquid ramp: (a) Three dimensional reconstruction. The height difference between three subsequent black lines is 1.0 mm. (b) Averaged height profiles for forcing to the l.h.s (black) and to the r.h.s. (gray). The solid lines represent fits by Eq.\,(\ref{eq:transport:hfit}) within the range between $x=60\,\mathrm{mm}$ and $x=120\,\mathrm{mm}$.} \label{fig:transport.ramp.3d}
\end{center}
\end{figure}

Figure~\ref{fig:transport.ramp.3d}\,(a) shows the surface topography of the magnetic liquid as reconstructed from the X-ray absorption, for a driving velocity of $v_G=4.1\,\frac{\mathrm{cm}}{\mathrm{s}}$ to the l.h.s. We averaged over $1300$ subsequent images with an exposure time of $0.4\,\mathrm{s}$ each. Obviously the liquid builds up a ramp in the container. A longitudinal cut of the ramp is presented in Fig.\,\ref{fig:transport.ramp.3d}\,(b) by black crosses. Each data point stems from an average along the y-dimension selecting the innermost $36.4\,\mathrm{mm}$ of the vessel. The gray symbols give the longitudinal cut for a driving to the r.h.s. The liquid ramp can be approximated by
\begin{equation}
h(x)=\tan{\left( \gamma \right)}x+h_0
\label{eq:transport:hfit}.
\end{equation}
as indicated by the fitted lines. The difference of the inclinations is measured by the angle $2\gamma$. This value is representative for the pressure gradient created by the magnetic pump.

\section{An Estimate of the Pressure Gradient}
The build up of the liquid ramp is caused by the Kelvin force, since we apply an inhomogeneous magnetic field to magnetized matter. The system can be described by the ferrohydrodynamic Bernoulli-equation \cite{rosensweig1985}
\begin{equation}
\rho \frac{\partial \vec{v}}{\partial t} + \vec{v} \cdot \vec{\nabla} \vec{v}=- \vec{\nabla} p + \mu_0 M \vec{\nabla} H + \eta \vec{\nabla}^2 \vec{v} + \rho \vec{g}.
\end{equation}
As a model assumption, we assume our system to contain no flow, thus all the terms with $v$ vanish. Further, we use a one-dimensional ansatz, and get
\begin{equation}
\frac{\partial p}{\partial x} = \mu_0 M_{\mathrm{eff}} \frac{\partial H_G}{\partial x}.
\label{eq:bernoulli_simp}
\end{equation}
The strength $H_G$ of the external magnetic field in harmonic approximation is given by Eq.\,(\ref{eq:modulation}). The effective magnetization is defined as
\begin{equation}
 M_{\mathrm{eff}}(x)=M \frac{h(x)}{h_0},
\label{eq:m_eff}
\end{equation}
where $h_0$ denotes the average height of the ferrofluid layer, and $M$ is assumed to be constant. Assuming a harmonic wave for the height modulation
\begin{equation}
 h(x)=h_0+\Delta h \sin{\left( \omega t -k_G x +\phi \right)},
\end{equation}
where $\phi$ denotes the phase difference with respect to the magnetic field, one gets
\begin{align}
M_{\mathrm{eff}}(x)&=M_{\mathrm{eff,ip}} \sin{(\omega t - k_G x)}  \nonumber \\
&- M_{\mathrm{eff,op}} \cos{(\omega t - k_G x)} + M_{\mathrm{eff},0}.
\label{eq:magnetisierung}
\end{align}
Here $M_{\mathrm{eff,ip}}$ denotes the part of the magnetization which is in phase with the external field according to Eq.\,(\ref{eq:modulation}). In contrast, $M_{\mathrm{eff,op}}$ is out of phase with $H_G$, but in phase with the gradient
\begin{equation}
 \frac{\partial H_G(x,t)}{\partial x} = -k \Delta H_G \cos{(\omega t - k_G x)}
\label{eq:feldgradient}
\end{equation}
of the external field. Inserting $M_{\mathrm{eff}}$ from Eq.\,(\ref{eq:magnetisierung}) into Eq.\,(\ref{eq:bernoulli_simp}) and averaging over one period of the external field leads to a mean pressure gradient
\begin{equation}
\left< \frac{\partial p}{\partial x} \right>_x = \frac{\int_0^{\lambda} \mu_0 M_{\mathrm{eff}} \frac{\partial H_G}{\partial x} \mathrm{dx} }{\lambda}
\label{eq:druckgradient}
\end{equation}
along the $x$ direction. This pressure gradient is experimentally determined with the observed inclination $\gamma$ of the surface:
\begin{equation}
\left< \frac{\partial p}{\partial x} \right>_x = \rho g \tan{\gamma}.
\label{eq:beta_von_druckgradient}
\end{equation}

\section{Experimental Results}
\begin{figure}
\centering
\includegraphics[width=0.8\linewidth]{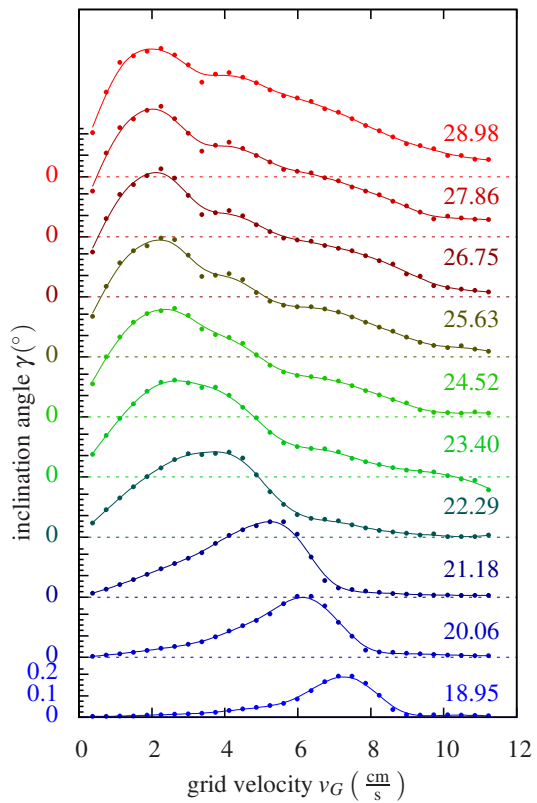}
\caption{Selection of surface inclinations versus the driving frequency for different $\mu_0 H_0$ (denoted by the right ordinate in mT), the curves are shifted equidistantly upwards to avoid intersections. The data points are connected by splines to guide the eye.} \label{fig:resonance_curves}
\end{figure}

In order to characterize the pressure difference created by the pump, we investigate its dependence on the externally applied field $H_0$ and the belt velocity $v_G$. Figure~\ref{fig:resonance_curves} presents the results for ten different values of $H_0$. For each field strength, $v_G$ was raised from $0$ to $11.2\,\frac{\mathrm{cm}}{\mathrm{s}}$ in $30$ steps.

Although the width and the shape of the $\gamma(v)$-curves changes with the applied field, in each case the pressure difference exhibits a maximum at a certain driving velocity $v_m(H_0)$, e.g.\,for $\mu_0 H_0=18.95\,\mathrm{mT}$ $\gamma$ reaches a maximum at $7.29\,\frac{\mathrm{cm}}{\mathrm{s}}$. With increasing $H_0$ this maximum is shifted to lower velocities. In order to determine the position of the maxima ($v_m$), we calculate the barycenter of those data being larger than $75\%$ of the maximal value of $\gamma$.

\begin{figure}
\centering
\includegraphics[scale=0.67]{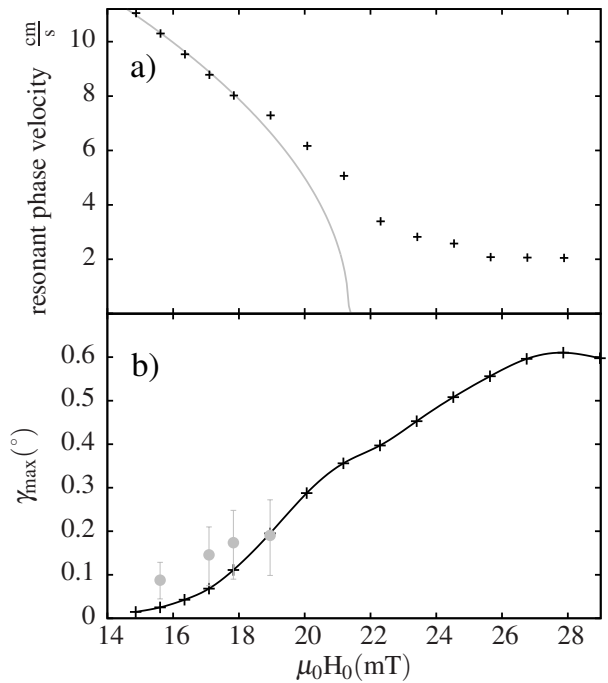}
\caption{a) Resonant phase velocity versus applied magnetic field. The black crosses give the experimental results. The gray line stems from Eq.\,(\ref{eq:transport.vphase}); b) Angles of maximal surface inclinations over the applied magnetic field. Black symbols are experimental data points. The black line just connects the points via a spline fit to guide the eyes. The gray circles are calculated according to Eq.\,(\ref{eq:magnetischer_druckunterschied}).}
\label{fig:max_angle}
\end{figure}

In Fig.\,\ref{fig:max_angle}\,a) the outcome is marked by black crosses. The gray curve denotes a model presented in the next section. The maximal surface inclination is extracted by fitting parabolas to the upper part of the curves. The result is plotted in Fig.\,\ref{fig:max_angle}\,b). It demonstrates a monotonic increase of the efficiency of the pump with the applied magnetic field, which seems to saturate at about $30\,\mathrm{mT}$.

\section{Modeling the Resonance}
The resonance of $\gamma$ for a specific $v_m$, as depicted in Fig.\,\ref{fig:resonance_curves}, resembles the resonance of the amplitudes of surface waves below the onset of the Rosensweig instability\cite{beetz2008,bashtovoi1977ews}. The resonant phase velocity $v_p(H_0)$ can be captured by the dispersion relation for an inviscid magnetic layer of infinite depth \cite{beetz2008}.

According to Eq.\,(\ref{eq:druckgradient}), we expect a monotonic relation between the wave amplitudes and the inclination $\gamma$. The function $v_m(H_0)$ in Fig.\,\ref{fig:max_angle}\,a) can be modeled in a certain regime by utilizing the dispersion relation for small surface wave amplitudes
\begin{equation}
\omega^2(k)= g_\mathrm{e} k+\frac{\sigma}{\rho}\,
k^3-\frac{\mu_0}{\rho}\frac{r}{r+1}\,k^2\, M^2.
\label{eq:transport.nonlin.dispersion.relation.short}
\end{equation}
It can be deduced from Eq.\,(36) in Ref.\,\cite{zelazo1969}, as detailed in Ref.\,\cite{reimann2005}. The effective permeability $r$ is the geometric mean $r=\sqrt{(1+\chi_\mathrm{ta})(1+\chi_\mathrm{ch})}$, where $\chi_\mathrm{ta}=(\partial M/\partial H)$ denotes the tangential susceptibility, and $\chi_\mathrm{ch}=(M/H)$ the chord susceptibility \cite{rosensweig1985}. For the resonant phase velocity at $k=k_G$ one yields
\begin{equation}
v_p=\frac{\omega}{k_G}=\sqrt{\frac{g_\mathrm{e}}{k_G}+\frac{\sigma}{\rho}\,k_G-\frac{\mu_0}{\rho} \frac{r}{r+1}\,M^2}.
\label{eq:transport.vphase}
\end{equation}
Using the jump condition for the field at the bottom of the container $H_{\mathrm{ext}} = H_i+M(H_i)$ and assuming that $H_{\mathrm{ext}}$ is approximately given by $H_0$, one can obtain $v_p(H_0)$ numerically. The outcome is marked in Fig.\,\ref{fig:max_angle}\,a) by the solid line. It has been calculated using the magnetization curve as presented in Fig.\,\ref{fig:magnetization.curve}. It shows a good agreement with the measured data only for small values of the field, because it is based on the assumption of small amplitudes of the surface waves. The intersection of this line with the x-axis at $\mu_0 H_{\mathrm{c,calc}}=21.35\,\mathrm{mT}$ denotes the estimate for the onset of the Rosensweig instability, where surface deformations spontaneously form without any external modulation. This critical value has been independently measured (see sec.\,\ref{sec:procedure} and Tab.\,\ref{tab:EMG909}) and matches within the experimental uncertainty.

The maximum of the pressure gradient has experimentally been observed for driving at $v_m$ as shown in Fig.\,\ref{fig:max_angle}\,b). The phase shift $\phi$ between $H_G$ and $M_{\mathrm{eff}}$ is assumed to be $\frac{\pi}{2}$ for this resonant driving, in analogy to a driven harmonic oscillator. Using Eqs.\,(\ref{eq:druckgradient}) and (\ref{eq:beta_von_druckgradient}) this assumption leads to
\begin{align}
\gamma=\frac{1}{\rho g} \left< \frac{\partial p}{\partial x} \right>_x &= \frac{\mu_0 M \, \Delta H_G \, k \, \Delta h}{2 \rho g h_0}.
\label{eq:magnetischer_druckunterschied}
\end{align}

We measured $\Delta h$ with a laser reflection method, as described in Ref.\,\cite{beetz2008}. When using this method, we assume a sinusoidal modulation of the surface, it is thus restricted to small values of the magnetic field $H_0$. Therefore we are restricted to values well below $H_{c,\mathrm{nonlin}}$. The corresponding data points for four measured values of $\Delta h$ are indicated in Fig.\,\ref{fig:max_angle}\,b) by circles. The data only agree within about $50 \%$. This is partly due to i) the flutter in $\lambda_G$, ii) the geometrical uncertainty connected with the laser reflection method, iii) the variation in $h_0$ along the ramp, iv) the variation of the magnetization and the field modulation within the layer thickness and v) the neglecting of the higher harmonics of $h(x,t)$. All these uncertainties are taken into account by the error bars. Moreover the crudeness of the model should not be brushed under the carpet. In particular it does not take into account any backflow effects and assumes a strictly two dimensional geometry.

\begin{table}
\centering \caption{Parameters measured for the ferrofluid, \emg Lot~H030308A from Ferrotec Co.}
\begin{tabular}{llll}
\hline\hline \vspace{0.1 ex}
density                                       & $\rho$                                          &    &  $0.9945\  $\,$\mathrm{g/cm}^3$     \\
surface tension                               & $\sigma$                                        &    &  $23.37 \  $\,$\mathrm{mN/m}$       \\
initial susceptibility                        & $\chi_0$                                        &    &  $0.95  \  $                        \\
viscosity                                     & $\eta$                                          &    &  $3.3   \  $\,$\mathrm{mPa\,s}$     \\
critical wavelength                           & $\lambda_\mathrm{c}=2\pi\,\sqrt{\frac{\sigma}{\rho g_0}}$                           &    &  $9.89\ $\,$\mathrm{mm}$  \\
calculated critical field                 & $\mu_0 H_\mathrm{c,calc}$                           &    &  $21.35\ $\,$\mathrm{mT}$\\
\hline
measured critical field                   & $\mu_0 H_\mathrm{c,meas}$                                &    &  $21\ $\,$\mathrm{mT}$ \\
\hline\hline
\end{tabular}
\label{tab:EMG909}
\end{table}

\section{Conclusion}
In this paper we have demonstrated a new kind of pump for magnetic fluids which is based on a resonance phenomenon in the advent of the Rosensweig instability. Our driving relies fundamentally on the excitation of surface waves. It is thus restricted to open channel geometries, and differs substantially from ``ferrohydrodynamic pumping'' as described in Ref.\,\cite{mao2006} which works in closed channels. That method, however, is most effective for frequencies comparable to the inverse Brownian relaxation time ($\approx 10\,\mathrm{kHz}$), while our driving mechanism favours low frequencies ($\approx10\,\mathrm{Hz}$).

We have characterized the pump under no-load conditions by measuring the static pressure gradient via a radioscopic method. Under variation of the driving velocity, the pressure gradient shows a resonance. The resonant phase velocity decreases with the applied magnetic field. This can be understood by calculating the phase velocity of surface waves from the dispersion relation, taking into account the nonlinear magnetization curve. The pressure gradient is observed to increase with the field, which is captured by a two dimensional model derived from the ferrohydrodynamic Bernoulli equation assuming a static equilibrium. This ansatz leads to a transport effect, whose strength grows linearly with the surface modulation, the peristaltic pumping on the other hand grows with its square. A refined model should include both effects and would be a task for numerical investigations.

In future work, the spatio-temporal driving of the liquid could be achieved in different ways. While in the present study we used a moving belt, harboring iron rods, to drive the liquid, alternatively the driving may be realized by a network of current carrying wires without any moving parts. This would allow to control the modulation amplitude and the bias of the magnetic field independently.

The pump has been tested under no-load conditions. The opposite approach would be to characterize the pump under maximum-load conditions. Therefore, the input and output of the pump must be connected by an open duct with minimal flow resistance, similar to the setup described in Ref.\,\cite{kikura1990}.

Note that our resonant driving may be exploited as well to drive non-magnetic liquids via the magnetically driven peristaltic motion investigated in Ref.\,\cite{park1999dmf}.

So far we have focused our measurements on the resonant behavior below the formation of surface undulations. It remains to be investigated how the spatio-temporal driving interacts with the secondary instabilities leading to more complicated surface patterns \cite{ruediger2007}.

\section{Acknowledgements}
The authors are grateful to A. Beetz and C. Groh for their contributions to the experimental design and K. Oetter for the mechanical realization. Further we want to thank K. Zimmermann and A. Naletova for discussion and DFG SFB 481 for financial support.


%

\end{document}